# New tissue engineered scaffolds for rotator cuff tendon-bone interface regeneration

Ting Zhang[1,#], Jianying Zhang[1,*]

1 Department of Orthopaedic Surgery, University of Pittsburgh, Pittsburgh, PA 15213, USA; zhangtin@pitt.edu (T.Z)

**\*Correspondence:**

**Jianying Zhang, PhD**

Department of Orthopaedic Surgery

University of Pittsburgh

E1640 BST, 200 Lothrop Street

Pittsburgh, PA15213, USA

Email: jianying@pitt.edu (J.Y.Z)

**Abstract**

Healing of Tendon-bone interface(TBI) injuries is slow and is often repaired with scar tissue formation that compromises normal function. Despite the increasing maturity of surgical techniques, re-tearing of the rotator cuff after surgery remains common. The main reason for this issue is that the original structure of the rotator cuff at the TBI area is difficult to fully restore after surgery, and anatomical healing of the rotator cuff TBI is challenging to achieve solely through surgery. With the advancement of tissue engineering technology, more and more basic researchers and clinical surgeons are recognizing the enormous potential of tissue engineering in promoting TBI healing. Growing research evidence indicates that tissue engineering technology not only effectively promotes repairing and remodeling of the TBI but also reduces the formation of fibrous vascular scar tissue, leading to more orderly tissue reconstruction. The core of tissue engineering technology approaches lies in combining the use of various scaffolds, cells and bioactive molecules to simulate the natural environment of TBI healing, achieving optimal therapeutic outcomes. In this review, we will systematically summarize and highlight recent progress in the application of tissue engineering on TBI regeneration, particularly focusing on advancements in novel scaffolds and their role and potential in promoting healing of the TBI of rotator cuff, providing valuable references for clinical application and research.

**Keywords:** tissue engineered scaffold; tendon-bone interface; tissue engineering; regeneration

**Introduction**

Rotator cuff injuries are very common, with a prevalence of approximately 30% in individuals over the age of 65[1]. The rotator cuff is one of the key structures responsible for maintaining the stability of the shoulder joint[2]. Studies have demonstrated that the rate of TBI healing after rotator cuff repair is suboptimal, with re-tear rates reaching as high as 20% to 94% [3–5]. The primary reason is that the rotator cuff structure cannot be fully restored to its original complex four-layer structure after rotator cuff repair [6]. The TBI consists of four layers: tendon, uncalcified fibrocartilage, calcified fibrocartilage, and bone, with complex interactions between these layers. In the normal four-layer structure, fibers and cells are arranged in a gradient ,orderly manner, and the matrix components exhibit gradient variations [7]. Current surgical methods can only restore the anatomical position of the tendon and fill the TBI with scar tissue, but this type of healing cannot achieve the original structure's stress-dispersing effect [8].

The healing procedures at the TBI involve a number of cell types and biological processes **[9]**. A promising approach among the various options is stem cell-based therapy, which leverages the cells' capacity for self-renewal and differentiation. This method facilitates tissue regeneration by activating either intrinsic or external regulatory pathways **[10,11]**. Several of our studies have demonstrated the role of kartogenin (KGN) and PRP in promoting tendon bone healing **[12－14]**. Studies have reported that performing bone marrow stimulation or footprint decortication at the rotator cuff insertion site during rotator cuff repair can induce the release of beneficial cytokines at the tendon－bone interface, thereby promoting tendon-to-bone healing and reducing the incidence of postoperative rotator cuff retears**[15]**   . Tissue engineering has demonstrated great advantages in promoting TBI healing, The scaffold is a crucial part of tissue engineering. The scaffold can flexibly combine with cells, bioactive molecules, and physical stimuli to form tissue-like structures that replace damaged tendon-bone tissues. Additionally, the scaffold can gradually release bioactive molecules, growth factors, and cells in vivo, promoting the proliferation and differentiation of osteoblasts and fibroblasts, thus providing continuous nutrients for tendon-bone healing. By mimicking the mechanical gradient environment of TBI, the scaffold achieves optimized mechanical distribution, reduces the risk of stress concentration, and thereby promotes healing and decreasing the rate of re-rupture **[16]**. This review will summarize and discuss recent advances in novel biomaterials used as tissue engineered scaffolds, supporting research and application in rotator cuff tendon-bone healing.

**1 Characteristics of tissue engineered scaffolds**

For scaffolds to promote tendon-bone healing, the material's mechanical properties, physical characteristics, and biocompatibility should also be considered, as these key traits determine clinical applicability and performance [17]. Because scaffolds are subjected to dynamic stresses in vivo, their mechanical properties directly impact their elasticity and longevity [18]. The physical characteristics of scaffolds, including structure and texture, are essential for stem cell attachment to the scaffold [19]. Compartmentalization defines the spatial configuration of scaffolds. The toughness and porosity of compartmentalized structures determine their energy absorption capacity and permeability to fluids and nutrients. The surface roughness of scaffolds can influence cellular response and tissue integration. The shape and size of scaffolds need to be customized to mimic the normal structure of tissues, thereby ensuring normal cellular activity. The biocompatibility of scaffolds reflects the material's integration properties without causing negative host reactions [20]. Additionally, the pH value, charge, and hydrophilicity of scaffolds can significantly impact protein adsorption, cell adhesion, and overall bioactivity. The functional groups and chemical structure of scaffolds can be designed to facilitate interaction spaces with cells and biomolecules [21].

The development of novel scaffolds for promoting tendon-bone healing has led to various types of scaffolds. Scaffolds can be categorized by material source into biologically derived scaffolds, inorganic scaffolds, and organic synthetic scaffolds [21].Biologically derived scaffold materials mainly originate from natural proteins (e.g., collagen) [22] or polysaccharides (e.g., chitosan) [23].These materials have good biocompatibility and can effectively promote cell attachment and proliferation, but they have lower mechanical strength and are easily degradable. Inorganic scaffold materials are mostly calcium phosphate-based, such as hydroxyapatite (HA) and tricalcium phosphate. They possess high mechanical properties and osteoconductivity [24], making them suitable for bone tissue repair. However, their degradation rate is difficult to regulate, and their bioactivity is limited. Organic synthetic scaffold materials are mainly composed of synthetic polymers, such as polylactic acid (PLA) [25],PLA-glycolic acid copolymer (PLGA) [26], and polyvinyl alcohol (PVA) [27]. Their mechanical properties and degradation rates are adjustable, making them suitable for various tissue engineering applications, though their biocompatibility is inferior to natural materials, necessitating surface modifications to enhance cell interactions. In addition to being classified by material source, novel scaffolds can be divided based on structural characteristics and functional features into single-layer scaffolds, multilayer integrated scaffolds, and gradient scaffolds, each with unique advantages and disadvantages, which are detailed below.

**2 Single-layer Scaffold**

Research on single-layer scaffolds started the earliest, with the most extensive studies, and they are the most commercially available type of scaffolds, such as Regeneten [28] and TissueMend [29]. Early single-layer scaffold materials mainly used degradable materials such as gelatin, focusing on releasing nutrients upon scaffold degradation as the therapeutic strategy. Later research showed that polycaprolactone (PCL) and PLGA materials have good biocompatibility and mechanical properties, gradually becoming common materials for constructing single-layer scaffolds. In terms of scaffold biomolecules, early research promoted cell growth and differentiation in specific regions of tendon-bone injury by regulating the secretion and spatial distribution of bioactive factors of the transforming growth factor-beta (TGF-β) and bone morphogenetic protein (BMP) superfamily [30–32]. Recent research has found that exosomes are nanoscale extracellular vesicles (EVs) with a lipid bilayer and membrane structure, serving as an effective platform for drug delivery in tendon-bone healing [33]. Tissue-engineering-aided exosomes can be stably and continuously delivered, promoting the regeneration and repair of TBI both functionally and structurally [34]. Gelatin Methacryloyl (GelMA) is a promising biomaterial with good biocompatibility and tunable mechanical properties, suitable for tissue engineering scaffolds and drug carriers [35].KGN-loaded GelMA hydrogel scaffolds were obtained through UV cross-linking and vacuum freeze-drying. Chenglong Huang et al. [35] demonstrated in a New Zealand rabbit rotator cuff repair study that microfracture combined with KGN-loaded GelMA hydrogel scaffolds enhances the repair of rotator cuff injuries by promoting fibrocartilage formation and improving the mechanical properties of the TBI [27].

Overall, although research on single-layer scaffolds is relatively mature in various aspects, the exploration and development of new scaffold materials and biomolecules are still ongoing. The research findings on single-layer scaffolds can guide the construction of multilayer scaffolds. The design of single-layer scaffolds is simple, focusing primarily on improving the microenvironment of the tendon-bone healing site, enriching seed or differentiated cells at the repair site, providing the necessary nutrients, and regulating their growth and differentiation, thereby promoting and optimizing the deposition of extracellular matrix to facilitate tendon-bone healing. Additionally, single-layer scaffolds possess good homogeneity and integrity, with strong shape plasticity. Since the shape of the tendon-bone junction after tendon-to-bone reconstruction is often variable and irregular, single-layer scaffolds' flexibility to match the repair site's shape is an important advantage.

## 3 Multilayer Integrated Scaffold

Studies have shown that the metabolism and proliferation of chondrocytes and osteoblasts require different microenvironments [36]. The single-layer scaffold structure contains only a single type of bioactive molecule, making it difficult to simultaneously promote stem cell differentiation into chondrocytes, osteoblasts, or fibroblasts. Therefore, multilayer integrated scaffolds were developed to address this limitation. Multilayer integrated scaffolds have multiple layers with different functional structures, allowing them to providing appropriate microenvironments for cells at both ends of the interface. Their design concepts are diverse, some scaffolds provide space for cell attachment and growth, bridging the defect on both sides of the reconstruction site through filling or covering while enriching repair cells as a treatment strategy. Common biodegradable synthetic materials used in tissue engineering include poly(L-lactic acid) (PLLA) [37], polyglycolic acid (PGA) [38], PLA-PLGA [39,40], and PCL [41]. Additionally, common natural polymers include alginate [42], silk [43], and collagen [44]. Scaffold combinations include PLLA-HA [45], PLGA-PCL-HA [46], collagen-HA [47], PCL-TCP-β [48], and CS-g-PCL-TGF-β3 [49], among others. Compared to single-layer scaffolds, multilayer integrated scaffolds have a more complete mechanism for promoting tendon-bone healing. Related research has confirmed their effectiveness in promoting tendon-bone healing [50,51].

## 4 Gradient Biomimetic Scaffold

Although multilayer integrated scaffolds can simulate the native TBI microenvironment through functional divisions of different layers, the transitions between native TBI layers are more gradual than abrupt. Therefore, gradient-transition scaffold materials have gradually been developed. These scaffolds simulate the natural gradient of normal TBI, providing an optimized environment for cellular activities, which promotes tendon-bone healing. Common technologies used to fabricate gradient biomimetic scaffolds include electrospinning, 3D printing, freeze-drying, melt pressing, and salt immersion techniques [52]. Electrospinning technology stands out among many nanofiber manufacturing methods due to its significant advantages, including the ability to rapidly produce a continuous fiber network and its versatility with various materials [53]. Yu et al. used photothermal welding technology on aligned fiber scaffolds to successfully create a gradient from alignment to randomness [54]. They premixed the photothermal material indocyanine green (ICG) directly into the polyurethane (PU) solution and used a drum collector for electrospinning to generate uniaxially aligned nanofibers. After fabrication, laser irradiation of ICG generates heat, causing the nanofibers to melt and weld at their melting points. Yang et al. [55] used a metal-organic framework (MOF) as a carrier and employed continuous electrospinning to match the longitudinal spatial morphology of multiple tissues, producing a biphasic metallic flexible electrospun fiber membrane [55]. MOF as a carrier not only allows for the sustained release of metal ions but also promotes the differentiation of stem cells on the scaffold into osteoblasts and fibroblasts. Gao et al. synthesized nanoparticles (Sr-MBG) using a sol-gel method and prepared biphasic inductive and immunoregulatory electrospun fiber scaffolds (BIIEFS) containing Sr-MBG. Experimental results showed that the electrospun scaffolds increased the number of M2 macrophages, while also observing synchronous regeneration of tendons, fibrocartilage, and bone, significantly enhancing the biomechanical strength of the supraspinatus tendon-humerus [56]. Kim et al. proposed using bioprinting to effectively distribute human adipose-derived stem cells (hASCs) and their biocomponents for TBI interface regeneration. Bionic ink derived from decellularized extracellular matrix (dECM) of porcine bone and tendon was used, with HA supplemented in the bone region and TGF-β/PVA supplemented in the tendon region [57].

Advanced manufacturing technologies provide more possibilities for preparing gradient artificial bone repair scaffolds. Currently, gradient electrospun fiber scaffolds have been developed with spatially differentiated fiber arrangements and bioactive substances, along with gradient 3D-printed scaffolds fabricated using layered stacking, graded porosity, and bio-3D printing technologies [57,58]. These scaffolds showed good biocompatibility in vitro and accelerated tissue regeneration in animal trials, with significant histological improvements observed[59]. However, preparing such scaffolds is challenging, and the gradient components, structures, and phenotypes of the scaffolds still require further exploration [59]. Moreover, simplifying the preparation process requires further investigation [58].

## 5 Summary and Outlook

Tissue engineered scaffolds have made significant progress in promoting tendon-bone healing, with the development of various types of scaffolds to address the challenges of the complex biological healing process. Single-layer scaffolds provide basic initial cell adhesion and support, but they lack the characteristics necessary to mimic the complexity of TBI, limiting their long-term efficacy. In contrast, multilayer integrated scaffolds, especially gradient biomimetic scaffolds, represent a significant leap in biomaterial design. By accurately replicating the natural gradient in TBI, these scaffolds provide a seamless gradient structure, improved load distribution, and enhanced growth and differentiation of stem cells. Artificial synthetic scaffolds that effectively promote tendon-bone healing have been designed and fabricated. However, issues persist, such as gradient scale mismatches, unclear material-tissue interactions, and side effects from degradation products. Therefore, future research should continue to explore, design, and fabricate synthetic scaffolds, investigating more cellular molecules, further optimizing the geometric structure and surface curvature of the scaffold to better simulate the physiological TBI structure, and synergistically guiding and regulating the behavior of differentiated stem cells to ultimately promote effective structural repair of TBI.

Although novel biomaterials have made some progress in promoting tendon-bone healing, there is still considerable potential for innovation and many challenges ahead. By focusing more on novel biomaterials, advanced manufacturing technologies, and interdisciplinary collaboration, we can further improve tissue engineering techniques, explore new methods for tendon-bone healing, and ultimately enhance patient outcomes.